\documentstyle[,epsfig]{l-aa}
\input psfig                  
\onecolumn

\begin{document}
   \hyphenation{brems-strah-lung
                Boltz-mann}
   \thesaurus{              
              02.04.2;  
              02.16.1;  
              02.18.5;  
              08.09.2;  
              08.09.2;  
              09.07.1;  
              10.03.1;  
              13.07.3   
              }
   \title{Positron propagation in semi-relativistic plasmas:
          particle spectra and the annihilation line shape
         }
   \author{I.~V.~Moskalenko\inst{1,2} \and E.~Jourdain\inst{1}}
   \offprints{I.V.~Moskalenko}
   \institute{Centre d'Etude Spatiale des Rayonnements (CNRS/UPS), 
              9 av.\ Colonel-Roche, 31028 Toulouse Cedex, France
   \and       present address:
              Max-Planck-Institut f\"ur Extraterrestrische Physik,
              Postfach 1603, D-85740 Garching, Germany
              }
   \date{Received ; accepted}

   \maketitle
   \markboth{I.V.~Moskalenko \& E.~Jourdain: Positron propagation in
             semi-relativistic plasmas
            }{}
   \begin{abstract}

By solving the Fokker-Planck equation directly we examine effects of
annihilation, particle escape and injection on the form of a
steady-state positron distribution in thermal hydrogen plasmas with
$kT<m_ec^2$. The positron fraction considered is small enough, so
it does not affect the electron distribution which remains Maxwellian.
We show that the escape of positrons in the form of electron-positron
pairs and/or pair plasma, e.g.\ due to the diffusion or radiation
pressure, has an effect on the positron distribution causing, in some
cases, a strong deviation from a Maxwellian. Meanwhile, the distortion
of the positron spectrum due to only annihilation is not higher than a
few percent and the annihilation line shape corresponds to that of
thermal plasmas. Additionally, we present accurate formulas in
the form of a simple expression or a one-fold integral for energy
exchange rates, and losses due to M\o ller and Bhabha scattering,
$e^+e^-$-, $ee$- and $ep$-bremsstrahlung in thermal plasmas as well as
due to Compton scattering in the Klein-Nishina regime.

Suggesting that annihilation features observed by {\it SIGMA} telescope
from Nova Muscae and the 1E~1740.7--2942 are due to the
positron/electron slowing down and annihilation in thermal plasma, the
electron number density and the size of the emitting regions have been
estimated. We show that in the case of Nova Muscae the observed
radiation is coming from a pair plasma stream ($n_{e^+}\approx
n_{e^-}$) rather than from a gas cloud. We argue that two models
are probably relevant to the 1E~1740.7--2942 source: annihilation in
(hydrogen) plasma $n_{e^+}\la n_{e^-}$ at rest, and annihilation in the
pair plasma stream, which involves matter from the source environment.

   \end{abstract}

\section {Introduction}
Positron production and annihilation are widespread processes in
nature. Gamma-ray spectra of many astrophysical sources exhibit an
annihilation feature, while their continuum indicates presence of
mid-relativistic thermal plasmas with $kT\la200-300$ keV. Spectra of
$\gamma$-ray bursts and Crab pulsar show emission features in the
vicinity of 400--500 keV (Mazets et al.\ 1982; Parlier et al.\ 1990),
which are generally believed to be red-shifted annihilation lines.
Recent observations with {\it SIGMA} telescope have exhibited
annihilation features in the vicinity of $\sim500$ keV in spectra of
two Galactic black hole candidates, 1E~1740.7--2942 (Bouchet et
al.\ 1991; Sunyaev et al.\ 1991; Churazov et al.\ 1993; Cordier et al.\
1993) and Nova Muscae (Goldwurm et al.\ 1992; Sunyaev et al.\ 1992). A
narrow annihilation line has been observed from solar flares (Murphy et
al.\ 1990) and from the direction of the Galactic center
(Leventhal et al.\ 1978).

The region of the Galactic center contains several sources which
demonstrate their activity at various wavelengths and particularly
above several hundred keV (e.g., see Churazov et al.\ 1994). Escape of
positrons from such a source or several sources into the interstellar
medium, where they slow down and annihilate, can account for the 511
keV narrow line observed from this direction. The 1E~1740.7--2942
object has been proposed as the most likely candidate to be responsible
for this variable source of positrons (Ramaty et al.\ 1992). This would
only require that a small fraction of $e^+e^-$-pairs, which is
generally believed to be produced in the hot inner region of an
accretion disc, escapes into surrounding space (Meirelles \& Liang
1993). Nova Muscae shows a spectrum which is consistent with
Comptonization by a thermal plasma $kT_e\la100$ keV in its hard
X-ray part, while a relatively narrow annihilation line observed by
{\it SIGMA} during the X-ray flare on 20--21 January, 1991 implies that
positrons annihilate in a much colder medium (Gilfanov et al.\ 1991;
Goldwurm et al.\ 1992).

Numerous studies of positron propagation and annihilation in cold
interstellar gas (e.g., see Guessoum et al.\ 1991 and references
therein) have been inspired by observations of a narrow 511 keV line
emission from the Galactic center region. Relativistic pair plasmas
have been a matter of investigation during a decade (for a review, see
Svensson 1990). In all thermal models, however, particles are assumed
to be Maxwellian {\it a priori} and very often one only pays attention
to the relevant relaxation time scales. Herewith, the annihilation line
shape, the main feature which can be actually measured, is
strongly influenced by the real particle distribution. The latter can
differ from a Maxwellian under certain circumstances, such as particle
injection, escape, and annihilation. It is thus of astrophysical
interest to study particle distributions in various physical
conditions.

In this paper we use a Fokker-Planck approach to examine the effects of
annihilation, particle escape and injection on the form of a
steady-state positron distribution in thermal hydrogen plasmas with
$kT<m_ec^2$. Pairs are assumed to be produced in the bulk of the plasma
due to $\gamma\gamma$-, $\gamma$-particle, or particle-particle
interactions, or to be permanently injected into the plasma volume by
an external source. We don't touch here upon the cause of particle
escape, it could be of diffusive origin or due to the radiation
pressure (e.g., see Kovner 1984). Since the plasma cloud serves as
a thermostat, it is therefore reasonable, as the first step into the
problem, to consider that the electron distribution approaches
Maxwellian. The positron fraction considered is small enough, so it
does not affect the electron distribution.

Suggesting that the features observed by {\it SIGMA} in $\ga300$ keV
region are due to the electron-positron annihilation in thermal
plasma, we apply the obtained results to Nova Muscae and
the 1E~1740.7--2942 source in order to get the parameters of the
emitting regions where the annihilation features have been observed.

In Sect.~2 the Fokker-Planck treatment is considered and we present a
method to obtain a steady-state solution. The reaction rate
formalism is introduced in Sect.~3. The expressions for energy changes
and losses due to Coulomb scattering, bremsstrahlung and Comptonization
are given in Sect.~4--6. Electron-positron annihilation is considered
in Sect.~7. The results of calculation are discussed in Sect.~8. In the
last section (Sect.~9) we discuss the physical parameters of the
emitting regions in Nova Muscae and the 1E~1740.7--2942 source.
Throughout the paper units $\hbar =c=m_e=1$ are used.

\section {The Fokker-Planck Equation: Positron Spectrum}
Assuming the isotropy of the positron energy distribution function
$f(\gamma)$, the steady-state Fokker-Planck equation takes the form
   \begin{equation}
{d\over d\gamma}\left\{{d\over d\gamma}[D(\gamma)f(\gamma)]
-P(\gamma)f(\gamma)\right\}
-[A(\gamma)+E(\gamma)]f(\gamma)+S(\gamma)=0,                \label{2.1}
   \end{equation}
where $\int f(\gamma)\,d\gamma=1$, $\gamma$ is the positron Lorentz
factor, $P(\gamma)\equiv d\gamma/dt$ is the dynamical friction (energy
loss rate), $D(\gamma)\equiv d(\Delta\gamma)^2/dt$ is the energy
dispersion rate, $A(\gamma)$ and $E(\gamma)$ are the
annihilation and the particle escape rates, respectively, and
$S(\gamma)$ is the positron injection term.

In the steady-state regime, without sources and sinks, the kinetic
coefficients obey the equation (Lifshitz \& Pitaevskii 1979) which
results from the absence of the flux density in the energy space,
   \begin{equation}
D'(\gamma)f_1(\gamma)+D(\gamma)f_1'(\gamma)=P(\gamma)f_1(\gamma),
                                                            \label{2.2}
   \end{equation}
where $f_1(\gamma)$ is to be a Maxwellian distribution
$f_1(\gamma)\sim\gamma(\gamma^2-1)^{1/2} e^{-\gamma /kT}$ ($kT$ is the
dimensionless plasma temperature). This equation fixes a relation
between the coefficients
   \begin{equation}
D(\gamma)={1\over f_1(\gamma)}\int_1^\gamma d\gamma'\,f_1(\gamma')P(\gamma').
                                                            \label{2.7}
   \end{equation}

We emphasize that the coefficients of the Fokker-Planck equation have
an additive property. They represent the sum of coefficients for various
processes which have to be evaluated separately.

Although the plasma cloud serves as a thermostat with true Maxwellian
distribution, annihilation and sinks distort the distribution
$f_1(\gamma)$. We are thus looking for the solution of eq.~(\ref{2.1})
in the form $f(\gamma)=f_1(\gamma)g(\gamma)$, which gives an equation
for the unknown function $g(\gamma)$ (Moskalenko 1995)
   \begin{equation}
g'(\gamma)=
{1\over D(\gamma) f_1(\gamma)}\left\{\int_1^\gamma (A+E)f
\,d\gamma'-\int_1^\infty (A+E)f\,d\gamma'\times\int_1^\gamma
\tilde S\,d\gamma'\right\},                                 \label{2.8}
   \end{equation}
Eliminating $g(\gamma)$ in favour of $f(\gamma)$ yields the
integro-differential equation for the distorted function
   \begin{equation}
f'(\gamma)-f(\gamma)\left\{{1\over\gamma}+{\gamma\over\gamma^2-1}
-{1\over kT}\right\}=  
{1\over D(\gamma)}\left\{\int_1^\gamma (A+E)f
\,d\gamma'-\int_1^\infty (A+E)f\,d\gamma'\times\int_1^\gamma
\tilde S\,d\gamma'\right\},                                 \label{2.3}
   \end{equation}
while $[Df]'-Pf=0$ at $\gamma=1$ was assumed (cf.\ eq.~(\ref{2.2})).
The last term in eq.~(\ref{2.3}) follows simply from conservation of
the total number of positrons
   \begin{equation}
\int_1^\infty[A(\gamma)+E(\gamma)]f(\gamma)\,d\gamma=\int_1^\infty
S(\gamma)\,d\gamma,                                         \label{2.4}
   \end{equation}
which is always fulfilled if the source function has the form
$S(\gamma)=\tilde S(\gamma)\times$ $\int_1^\infty
(A+E)f\,d\gamma'$ and $\int_1^\infty\tilde S\,d\gamma=1$. A
regular singular point $\gamma=1$ in eq.~(\ref{2.3}) does not lead to
any singularity of the solution, which is Maxwellian-like at the
low-energy part. Equation (\ref{2.4}) gives also an idea of physical meaning
of term $[E(\gamma)f(\gamma)]$, that is the number of positrons with
Lorentz factor $\gamma$ escaping from the plasma volume per 1 sec. The
approach can be easily generalized to include inelastic processes,
stochastic acceleration etc.

Equation (\ref{2.8}) or (\ref{2.3}) can be resolved numerically with an
algorithm which reduces it to a first-order differential equation. Let
$f_i(\gamma)$ is the solution obtained after the $i$-th iteration, then
the equation
   \begin{equation}
f'_{i+1}(\gamma)-f_{i+1}(\gamma)\left\{{1\over\gamma}
+{\gamma\over\gamma^2-1}-{1\over kT}\right\}=
{1\over D(\gamma)}\left\{\int_1^\gamma (A+E)f_i
\,d\gamma'-\int_1^\infty (A+E)f_i\,d\gamma'\times\int_1^\gamma
\tilde S\,d\gamma'\right\},                                 \label{2.5}
   \end{equation}
with the initial condition\footnote{
which follows from suggestion $f(\gamma)=f_1(\gamma)g(\gamma)$,
 where $f_1(\gamma)$ is a Maxwellian.
} 
$f_{i+1}(1)=0$ allows us to get the next approximation
$f_{i+1}(\gamma)$ of the solution. For eq.~(\ref{2.8}), the condition
$g_{i+1}(1)=g_i(1)$  could be taken. To start the iteration procedure
one can use the Maxwell-Boltzmann distribution $f_1(\gamma)$, although,
in some cases, when a solution of eq.~(\ref{2.5}) deviates strongly
from Maxwellian, that causes a deviation in normalization during
first iterations. Since a solution of eq.~(\ref{2.5}) $f_n(\gamma)$
multiplied by a constant would be also a solution, it has to be
normalized in the end of iteration process. This algorithm converges
quickly and gives a good approximation of the solution already after
several iterations. The actual signature of the convergence could be an
equality $\int_1^\infty (A+E)f_{i-1}\,d\gamma'= \int_1^\infty
(A+E)f_i\,d\gamma'$. The combination of functions
$(af_i+(1-a)f_{i-1})$, where $a=const\la1$, on the place of $f_i$ in
the right side allows sometimes to get a convergence faster.

\section{Reaction Rate Formalism}
Below we describe a formalism, which further allows us to calculate the
annihilation rate, energy losses and energy dispersion rate due to
Coulomb scattering, bremsstrahlung, and Comptonization.

The relativistic reaction rate $R$ for two interacting distributions of
particles is given by 
   \begin{equation}
R={1\over (1+\delta_{12})}\int dn_1\int dn_2\,\sigma(\vec\beta_1,\vec\beta_2)
(1-\vec\beta_1\cdot\vec\beta_2)\beta_r,                     \label{3.1}
   \end{equation}
where $\sigma(\vec\beta_1,\vec\beta_2)$ is the cross section of a
reaction, $dn_i$ and $\vec\beta_i$ are correspondingly the differential
number density and velocity of particles of type $i$ in the
laboratory system (LS), $\beta_r$ is the relative velocity of the
particles, the factor $(1+\delta_{12})^{-1}$ corrects for double
counting if the interacting particles are identical.

We consider energetic particles which interact with particles of a thermal
gas. Let masses of both types of particles be equal ($m_i=1$). For
isotropic distributions, eq.~(\ref{3.1}) can be reduced to the triple
integral over particle momenta, $p_i=\beta_i\gamma_i$,  and the relative
angle, $\cos\theta={\vec p_1\cdot\vec p_2/ p_1p_2}$,
   \begin{equation}
R={n_1n_2\over 2(1+\delta_{12})}\int_0^\infty dp_1\,{p_1^2\over
\gamma_1}f_1(p_1)\int_0^\infty dp_2\,{p_2^2\over\gamma_2}f_2(p_2)
\int_{-1}^1 d(\cos\theta)\,\gamma_r\beta_r\sigma(\gamma_r),
                                                            \label{3.2}
   \end{equation}
where $n_i$ is the number density of particles of type $i$ in the LS,
$f_i(p_i)$ are the momentum distribution functions,
$\int_0^\infty dp_i\,p_i^2f_i(p_i)=1$,
   \begin{equation}
\gamma_r=(1-\beta_r^2)^{-1/2}=\gamma_1\gamma_2
(1-\beta_1\beta_2\cos\theta)                                \label{3.3} 
   \end{equation}
is the relative Lorentz factor of two colliding particles (invariant).

Putting the relativistic Maxwell-Boltzmann distribution for the
electron gas (pay attention to the normalization) together with the
monoenergetic distribution for the beamed particles,
   \begin{equation} 
f_1(p_1)={e^{-\gamma_1/kT}\over kT K_2(1/kT)},              \label{3.4} 
   \end{equation}
   \begin{equation} 
f_2(p_2)={1\over p_2^2}\delta (p_2-p),                     \label{3.13}
   \end{equation}
into eq.~(\ref{3.2}) yields
   \begin{equation} 
R(\gamma)={n_1n_2\over 2(1+\delta_{12})kT K_2(1/kT)\gamma}
\int_{-1}^1 d(\cos\theta)\int_0^\infty dp_1\,{p_1^2\over\gamma_1}
\gamma_r\beta_r\sigma(\gamma_r) e^{-\gamma_1/kT},           \label{3.5} 
   \end{equation}
where $K_j$ is the $j$-order modified Bessel function.

Using eq.~(\ref{3.3}) to eliminate $\cos\theta$ in favor of
$p_r=\gamma_r\beta_r$ and changing variables from $p_1$ to $\gamma_1$
one can find
   \begin{equation} 
R(\gamma)={n_1n_2\over 2(1+\delta_{12})kT K_2(1/kT)\gamma^2\beta}
\int_0^\infty dp_r\int_{\gamma_1^-}^{\gamma_1^+}d\gamma_1\,
{p_r^2\over\gamma_r}\sigma(p_r)e^{-\gamma_1/kT},            \label{3.6}
   \end{equation}
where $\gamma_1^\pm = \gamma\gamma_r(1\pm\beta\beta_r)$. After
integrating over $\gamma_1$, the reaction rate can be exhibited in the
form (Dermer 1985)
   \begin{equation} 
R(\gamma)={n_1n_2\over(1+\delta_{12})K_2(1/kT)\gamma^2\beta}
\int_0^\infty dp_r\,{p_r^2\over\gamma_r}\sigma(p_r)
\sinh(\gamma\gamma_r\beta\beta_r/kT)e^{-\gamma\gamma_r/kT}. \label{3.7}
   \end{equation}

Another form of the reaction rate for interacting isotropic
distributions of particles (eqs.~[\ref{3.4}], [\ref{3.13}]) was found
useful for some purposes (Dermer 1984)
   \begin{equation}
R(\gamma)={n_1n_2e^{\gamma/kT}\over 2(1+\delta_{12})kTK_2(1/kT)
\gamma^2\beta}\int_1^\infty\! d\gamma_r
\int_{\gamma_c^-}^{\gamma_c^+}d\gamma_c\,\gamma_c\gamma_r\beta_r
\sigma(\gamma_r)\sqrt{2(\gamma_r+1)}
e^{-\gamma_c\sqrt{2(\gamma_r+1)}\over kT},                 \label{3.12}
   \end{equation}
where $\gamma_c={\gamma_1+\gamma_2\over\sqrt{2(\gamma_r+1)}}$ is the
Lorentz factor of the center-of-mass system (CMS), and
$\gamma_c^\pm={\gamma(1+\gamma_r\pm\beta
\gamma_r\beta_r)\over\sqrt{2(\gamma_r+1)}}$.

If we are interested in energy losses suffered by the energetic
particles in an isotropic gas, it is necessary to weight the cross
section in eq.~(\ref{3.6}) or (\ref{3.12}) by the average LS energy
change per collision $\langle\Delta\gamma\rangle$. The concrete form
for $\langle\Delta\gamma\rangle$ depends on the studied process.
Hereafter we will consider the reaction rate and energy losses per one
positron in the unit volume ($n_2=1$), while $n_e\equiv n_1$ will
denote the electron number density.

\section{Coulomb Collisions}
Speaking about the Coulomb scattering one usually implies the lowest
order approximation, which is called M\o ller scattering when referred
to identical particles $e^\pm e^\pm$, and Bhabha scattering
when referred to distinct particles $e^+e^-$. The effect of
bremsstrahlung in $ee$-collisions is strictly not separable from that
of scattering, however, it is convenient and generally accepted to
treat them separately.
Expressions for Coulomb energy losses and dispersion have been obtained
by Dermer (1985) and Dermer \& Liang (1989). Here we describe briefly
their results for the self-consistency of consideration.
 
The average LS energy change during a collision is (asterisk denotes
CMS variables)
   \begin{equation} 
\langle\Delta\gamma\rangle={1\over\sigma_{Coul}^*(\gamma_r)}
\int d^3p'^*{d^3\sigma_{Coul}^*\over dp'^{*3}}\Delta\gamma, \label{4.1} 
   \end{equation}
where $(d^3\sigma_{Coul}^*/dp'^{*3})$ is the differential cross
section, $d^3p'^*=p'^{*2}dp'^*d(\cos\psi'^*)d\phi'^*$, $\psi'^*$ and
$\phi'^*$ are the polar and azimuthal angles, respectively. The LS
energy change expressed in these variables is
   \begin{equation} 
\Delta\gamma=\gamma_c(\gamma'^*-\gamma^*)+\beta_c\gamma_c
[(p'^*\cos\psi'^*-p^*)\cos\omega^*-p'^*\sin\psi'^*cos\phi'^*\sin\omega^*],
                                                            \label{4.2} 
   \end{equation} 
where $\beta_c$ is the CMS velocity,
   \begin{equation}
   \begin{array}{l}
\gamma^*=\sqrt{(\gamma_r+1)/2},\\
p^*=\sqrt{(\gamma_r-1)/2},                                  \label{4.3} 
   \end{array}
   \end{equation}
are the Lorentz factor and momentum of a particle in the CMS prior to
scattering, $\gamma'^*$ and $p'^*$ are those after scattering, and
$\omega$ is a kinematic angle
   \begin{equation} 
   \begin{array}{l}
\cos\omega^*=(\vec\beta^*\cdot\vec\beta_c)/\beta^*\beta_c,\\
\sin\omega^*=\beta_1\beta_2\sin\theta/\gamma_r\beta_r\gamma_c\beta_c.
                                                            \label{3.9}
   \end{array}
   \end{equation}

Energy losses of a particle due to elastic Coulomb scattering are given
by eq.~(\ref{3.12}) with the cross section weighted by
$\langle\Delta\gamma\rangle$. Using azimuthal symmetry of the cross
section, Dermer (1985) obtains
   \begin{displaymath}
{d\gamma\over dt}= {n_e e^{\gamma/kT}\over
2(1+\delta_{12})kTK_2(1/kT)\gamma^2\beta}\int_1^\infty
d\gamma_r\,\beta_r\gamma_r\sqrt{2(\gamma_r+1)}\int d\gamma'^*\int
d(\cos\psi'^*){d^2\sigma_{Coul}^*\over d\gamma'^*d(\cos\psi'^*)} 
   \end{displaymath}
   \begin{equation} 
   \hskip 3cm
\times\int_{\gamma_c^-}^{\gamma_c^+}d\gamma_c\,e^{-\gamma_c
\sqrt{2(\gamma_r+1)}\over kT}\left\{\gamma_c(\gamma'^*-\gamma^*)+
({p'^*\over p^*}\cos\psi'^*-1)(\gamma-\gamma_c\gamma^*)\right\},
                                                            \label{4.4} 
   \end{equation}
since $\cos\omega^*=(\gamma-\gamma_c\gamma^*)/\gamma_c\beta_cp^*$. In
the case of elastic scattering $\gamma^*=\gamma'^*$ and $p^*=p'^*$,
that gives
   \begin{displaymath}
{d\gamma\over dt}= {n_e\over K_2(1/kT)\gamma^2\beta} \int_1^\infty
d\gamma_r\,\beta_r\gamma_r e^{-\gamma\gamma_r/kT}Y
   \end{displaymath}
   \begin{equation} 
   \hskip 3cm
\times\left\{\left(\gamma
p^{*2}+{\gamma^*kT\over\sqrt{2(\gamma_r+1)}}\right)\sinh
(\gamma\gamma_r\beta\beta_r/kT)-\gamma\beta p^*\gamma^*\cosh
(\gamma\gamma_r\beta\beta_r/kT)\right\},                    \label{4.5}
   \end{equation}
where 
   \begin{equation} 
Y=\int_{\psi'^*_{min}}^{\psi'^*_{max}} d(\cos\psi'^*)\,
(1-\cos\psi'^*){d\sigma_{Coul}^*\over d(\cos\psi'^*)}.      \label{4.6} 
   \end{equation}
The value of $\psi'^*_{max}$ can be assigned from geometrical
consideration: $\pi$ for distinct particles and $\pi /2$ for identical
particles. The minimum scattering angle $\psi'^*_{min}$ can be related
to the excitation of a plasmon of energy $\omega_p$. The correction for
double counting in the case of identical particles appears now as the
above condition for $\psi'^*_{max}$.

Integration of eq.~(\ref{4.6}) with M\o ller ($e^\pm e^\pm$) and Bhabha
($e^\pm e^\mp$) scattering cross sections (Jauch \& Rohrlich 1976)
gives
   \begin{equation} 
Y_{M\o }={2\pi r_e^2\over\gamma^{*2}(\gamma^{*2}-1)^2}
\left\{{1\over2}(2\gamma^{*2}-1)^2\left(\ln\Lambda+\ln\sqrt2+{1\over2}\right)
-\left(2\gamma^{*4}-\gamma^{*2}-{1\over4}\right)\ln2
+{1\over8}(\gamma^{*2}-1)^2\right\},                        \label{4.8}    \end{equation}
   \begin{equation} 
Y_{Bh}={2\pi r_e^2\over \gamma^{*2}(\gamma^{*2}-1)^2}
\left\{{1\over2}(2\gamma^{*2}-1)^2\ln\Lambda-{\beta^{*2}\over 24}
(22\gamma^{*4}+14\gamma^{*2}-\beta^{*2}+6)\right\}.         \label{4.9} 
   \end{equation}
The term $\ln\Lambda=\ln\sqrt{1-\cos\psi'^*_{max}\over
1-\cos\psi'^*_{min}}$ appearing in eqs.~(\ref{4.8})--(\ref{4.9}) is
the Coulomb logarithm. It is a slowly varying function of $\gamma^*$,
and often can be approximated by a constant. In the Born regime for the
cold plasma limit, the Coulomb logarithm is given by Dermer (1985)
$\ln\Lambda_{e^\pm e^\pm}=\ln\left({m_ec^2\over\hbar\omega_p}
(1-{1\over\gamma})\sqrt{\gamma+1}\right)$,
$\ln\Lambda_{e^\pm e^\mp}=\ln\Lambda_{e^\pm e^\pm}+\ln\sqrt2$.
Where the plasma frequency $\omega_p$ can be obtained from the usual
expression by replacing the electron rest mass with an average inertia
per gas particle $\langle\gamma\rangle_{kT}$ (Gould 1981),
$\omega_p^2=4\pi r_ec^2n_e/\langle\gamma\rangle_{kT}$.

Substitution of the Rutherford cross section yields the cold plasma
limit
   \begin{equation} 
{d\gamma\over dt}= -{4\pi r_e^2n_e\over\beta}\ln\Lambda.   \label{4.14}
   \end{equation}

The energy dispersion coefficients $d(\Delta\gamma)^2/dt$ can be
obtained from eq.~(\ref{2.7}). Another way is to square eq.~(\ref{4.2})
and to follow the above-described method. For M\o ller scattering of
an electron by a thermal electron distribution the correct form of the
coefficient has been obtained by Dermer \& Liang (1989)
   \begin{displaymath}
\left[{d(\Delta\gamma)^2\over dt}\right]_{M\o}=
{n_e e^{\gamma/kT}\over2kTK_2(1/kT)\gamma^2\beta}\int_1^\infty d\gamma_r
{(\gamma_r^2-1)\over\gamma^*\beta^*}
   \end{displaymath}
   \begin{equation}
   \hskip 3cm 
\times\left\{\eta_0\left({\cal I}_1\gamma^2-{{\cal I}_2\over2}
(\gamma^2+\gamma^{*2}\beta^{*2})\right)-\eta_1\gamma\gamma^*
\left(2{\cal I}_1-{\cal I}_2\right)+\eta_2\left({\cal I}_1\gamma^{*2}
-{{\cal I}_2\over2}\right)\right\},                        \label{4.15} 
   \end{equation}
where
   \begin{displaymath}
{\cal I}_1={2\pi r_e^2\over\gamma^{*2}(\gamma^{*2}-1)^2}\left[
{1\over2}(2\gamma^{*2}-1)^2+(1-2\ln2)\left(2\gamma^{*4}-\gamma^{*2}
-{1\over4}\right)+{1\over12}(\gamma^{*2}-1)^2\right],     
   \end{displaymath}
   \begin{displaymath}
{\cal I}_2={2\pi r_e^2\over\gamma^{*2}(\gamma^{*2}-1)^2}\left[
(2\gamma^{*2}-1)^2(\ln\Lambda+\ln\sqrt2)-\left(2\gamma^{*4}-\gamma^{*2}
-{1\over4}\right)+{1\over6}(\gamma^{*2}-1)^2\right],      
   \end{displaymath}
   \begin{displaymath}
\eta_i=\int_{\gamma_c^-}^{\gamma_c^+}d\gamma_c\,\gamma_c^i
e^{-\gamma_c\sqrt{2(\gamma_r+1)}\over kT}.                
   \end{displaymath}

\section{Bremsstrahlung}
Electron-positron bremsstrahlung is a well-known QED process, but the
calculation of its fully differential cross section for the photon
production is very laborious, the resulting cross section formula is
extremely lengthy and it was obtained quite recently (Haug 1985a,b). In
$e^+e^-$-collisions both particles radiate, and that brings some
uncertainties in calculation of the particle energy loss, increasing
particularly as the positron energy closes in the electron gas
temperature. The exact energy loss rate can be obtained using the cross
section differential in the energy of the outgoing positrons, but no
expression for this quantity is available. As it will be shown, the
bremsstrahlung energy loss is small in comparison with Coulomb and
Compton scattering losses, and that allows us to approximate it by the
radiated energy rate. We shall, hereafter, speak about the particle
energy loss taking into account the above remark.

An average energy loss through bremsstrahlung is given by
   \begin{equation} 
\langle\Delta\gamma\rangle=-{1\over\sigma_b(\gamma_r)}\int dk^*\int
d\Omega^*\,k{d^3\sigma_b^*\over dk^*d\Omega^*},             \label{5.1} 
   \end{equation}
where $(d^3\sigma_b^*/ dk^*d\Omega^*)$ is the bremsstrahlung
differential cross section in the CMS, and $k$ is the LS energy of the
radiated photon. It can be expressed as
   \begin{equation}
\langle\Delta\gamma\rangle={\gamma_cQ_{cm}\over\sigma_b(\gamma_r)},
                                                            \label{5.2} 
   \end{equation}
where $Q_{cm}(\gamma_r)=\int dk^*\,k^*({d\sigma_b^*/dk^*})$. For $e^+e^-$
bremsstrahlung Haug (1985c) gives an approximation
   \begin{equation} 
Q_{cm}={16\over3}\alpha r_e^2\left\{
\begin{array}{ll}
2\,(1.096-0.523p^*+0.1436p^{*2}+1.365p^{*3}-0.532p^{*4}), &\gamma^*\la8/5;\\
3\,(\gamma^*\ln(\gamma^*+p^*)-\gamma^*/6-0.726
+1.575\gamma^{*-1}-0.796\gamma^{*-2}), & \gamma^*\ga 8/5,
\end{array} \right.                                         \label{5.3} 
   \end{equation}
where $\alpha$ is the fine structure constant, and $p^*$, $\gamma^*$
are the CMS variables given by eq.~(\ref{4.3}).
The same for $e^\pm e^\pm$ bremsstrahlung is (Haug 1975)
   \begin{equation} 
Q_{cm}\simeq 8\alpha r_e^2{p^{*2}\over\gamma^*}\left\{
1-{4p^*\over 3\gamma^*}+{2\over3}\left(2+{p^{*2}\over\gamma^{*2}}\right)
\ln(\gamma^*+p^*)\right\}.                                  \label{5.4} 
   \end{equation}
Then, starting from eq.~(\ref{3.6}) and taking into account eq.~(\ref{5.2}) 
we get
   \begin{displaymath}
{d\gamma\over dt}={n_e\over\sqrt2 (1+\delta_{12})K_2(1/kT)\gamma^2\beta}
\int_0^\infty dp_r\,\beta_r(\gamma_r-1)^{1/2}Q_{cm}(\gamma_r)
e^{-\gamma\gamma_r/kT}
   \end{displaymath}
   \begin{equation}
   \hskip 3cm
\times\{\gamma\beta\gamma_r\beta_r
\cosh(\gamma\beta\gamma_r\beta_r/kT)-(kT+\gamma+\gamma\gamma_r) 
\sinh(\gamma\beta\gamma_r\beta_r/kT)\}.                     \label{5.5}
   \end{equation}

In a hydrogen plasma the moving positron suffers energy losses due to
$e^+e^-$- and $ep$-bremsstrahlung. For equal $e^-$ and $p$ densities,
$e^+e^-$ bremsstrahlung gives the dominant contribution to the energy
loss in the whole energy range.
At the high energy limit $e^+e^-$-bremsstrahlung energy loss becomes
equal to that of $ep$ and exactly twice the $ee$ energy loss; herewith
in the Born approximation $e^+p$ and $e^-p$ cases are identical (Jauch
\& Rohrlich, 1976). An expression for energy loss due to 
$ep$-bremsstrahlung was obtained by Stickforth (1961)
   \begin{equation} 
{d\gamma\over dt}= -{2\over3}n_e\alpha r_e^2
\left\{ \begin{array}{ll}
8\gamma\beta[1-(\gamma-1)/4
+0.44935(\gamma-1)^2-0.16577(\gamma-1)^3], & \gamma\la2; \\
\beta^{-1}[6\gamma\ln(2\gamma)-2\gamma-0.2900], & \gamma\ga2.
\end{array} \right.                                         \label{5.6} 
   \end{equation}

\section{Compton Scattering}
The presence of photons in a thermal plasma leads to essential energy
losses due to Compton scattering. Thomson limit remains a good
approximation while the photon energy is $\ll 1$ (the rest mass of
the electron) and the electron Lorentz factor is not too high. As the
photon energy reaches $\sim 0.1$ the difference from the classical
limit becomes large, the principal effect is to reduce the cross
section from its classical value. Numerous X-ray experiments show that
the actual temperature of plasmas in astrophysical sources (far)
exceeds 0.05 and the particle Lorentz factor exceeds often few units,
that is why we consider the Klein-Nishina cross section.

The particle energy loss rate due to Compton scattering is given by
   \begin{equation} 
{d\gamma\over dt}={1\over2\gamma^2\beta}
\int_0^\infty d\omega\,f_\gamma(\omega)\int_{k^-}^{k^+}dk\,k
\sigma_{KN}(k)\langle\Delta\gamma\rangle,                   \label{6.5} 
   \end{equation}
where
$\gamma$, $\beta$ are the LS particle Lorentz factor and speed prior to
scattering, $\omega$ is the initial photon energy in the LS, the
background photon distribution $f_\gamma(\omega)$ is normalized on the
photon number density $n_\gamma=\int d\omega\,\omega^2f_\gamma(\omega)$
or on the energy density as
$U_{ph}=\int\omega^3f_\gamma(\omega)\,d\omega$,
$k^\pm=\omega\gamma(1\pm\beta)$, and an average particle energy change
due to the scattering is
   \begin{equation} 
\langle\Delta\gamma\rangle={1\over\sigma_{KN}}\int dk'\int 
d(\cos\theta')\,{d^2\sigma_{KN}\over dk'd(\cos\theta')}\Delta\gamma.
                                                            \label{6.1} 
   \end{equation}
The Klein-Nishina differential cross section in the positron-rest-system
(PRS) is expressed in terms of initial $k$ and final $k'$ photon
energies (Jauch \& Rohrlich 1976),
   \begin{equation} 
{d\sigma_{KN}\over d(\cos\theta')}=\pi r_e^2\left({k'\over k}\right)^2
\left({k\over k'}+{k'\over k}-\sin^2\theta'\right),         \label{6.2} 
   \end{equation}
   \begin{displaymath} 
{k'\over k}={1\over 1+k(1-\cos\theta')},
   \end{displaymath}
where $\theta'$ is the photon scattering angle in this system. The
particle energy change in the LS due to the recoil effect is
   \begin{equation} 
\Delta\gamma=\omega-k'\gamma(1+\beta\cos\rho'\cos\theta'),  \label{6.3}
   \end{equation}
where $\rho'$ is the angle between the incoming photon and positron
velocity vectors in the PRS, $\beta\cos\rho'=(\omega/\gamma k)-1$ .

After the integration one can obtain
   \begin{equation} 
{d\gamma\over dt}={\pi r_e^2\over2\gamma^2\beta}\int_0^\infty 
d\omega\,f_\gamma(\omega)[S(\gamma,\omega,k^+)-S(\gamma,\omega,k^-)],
                                                            \label{6.6} 
   \end{equation}
where
   \begin{displaymath}
S(\gamma,\omega,k)=\omega
\left\{ \left(k+{31\over6}+{5\over k}+{3\over2k^2}\right)\ln(2k+1)
-{11\over6}k-{3\over k}+{1\over12(2k+1)}+{1\over12(2k+1)^2}
+Li_2(-2k)
\right\}
   \end{displaymath} 
   \begin{equation} 
   \hskip 5cm
-\gamma
\left\{ \left(k+6+{3\over k}\right)\ln(2k+1)
-{11\over6}k+{1\over4(2k+1)}-{1\over12(2k+1)^2}
+2Li_2(-2k)
\right\},                                                    \label{6.7}
   \end{equation}
and $Li_2$ is the dilogarithm
   \begin{displaymath}
Li_2(-2k)=-\!\int_0^{-2k}\!\!\!\ln(1-x){dx\over x}
=\left\{
\begin{array}{ll}
\sum_{i=1}^\infty(-2k)^i/i^2, & k\le0.2; \\
-1.6449341+{1\over2}\ln^2 (2k+1)-\ln (2k+1)\ln (2k)+\sum_{i=1}^\infty
i^{-2}(2k+1)^{-i}, & k\ge0.2.
\end{array} \right.
   \end{displaymath} 

Formulas (\ref{6.6})--(\ref{6.7}) give exactly the same result as
Jones' (1965) eq.~(13). The delta-function approximation of the photon
distribution $f_\gamma(\omega)\sim\delta(\omega-\omega_0)/\omega^2$ can
 sometimes be used for evaluation of the integral (\ref{6.6}). We
have found that in some cases it shows a good agreement with exact
calculations, e.g.\ for the Planck's distribution with $\omega_0=2.7kT$
(see Fig.~\ref{f2}).

The Thomson limit of the Compton scattering can be obtained similarly
by equating $k=k'$ in eq.~(\ref{6.2})
   \begin{equation}
\left({d\gamma\over dt}\right)_T=-{32\over9}\pi r_e^2 U_{ph}
\gamma^2\beta^2.                                            \label{6.8} 
   \end{equation}
For the energy dispersion rate one can get
   \begin{equation}
\left[{d(\Delta\gamma)^2\over dt}\right]_T={56\over45}\pi r_e^2
\langle\omega^2\rangle\gamma^2\beta^2(6\gamma^2\beta^2+1),
                                                            \label{6.9} 
   \end{equation}
where $\langle\omega^2\rangle=\int\omega^4
f_\gamma(\omega)\,d\omega$.

\section{Annihilation Rate and Spectrum}
The annihilation rate $A(\gamma)$ for monoenergetic positrons in
Maxwell-Boltz\-mann electron gas can be directly obtained from
eq.~(\ref{3.7}) by substitution of the annihilation cross section
(Jauch \& Rohrlich 1976)
   \begin{equation} 
\sigma_a(\gamma_r)={\pi r_e^2\over\gamma_r\beta_r^2(\gamma_r+1)}
\left\{ (\gamma_r+4+{1\over\gamma_r})\ln(\gamma_r+\sqrt{\gamma_r^2-1})
-\beta_r(\gamma_r+3)\right\}.                               \label{7.1}
   \end{equation}

The spectrum of photons $d\Gamma/d\nu$, which are emitted in the
annihilation is given by
   \begin{equation}
{d\Gamma\over d\nu}=n_+ n_- \int f_+(\gamma_+)\,d\gamma_+\int f_-(\gamma_-) 
H(\nu,\gamma_+,\gamma_-)\,d\gamma_- ,                       \label{7.2}
   \end{equation}
where $\nu$ is the dimensionless photon energy, $f_\pm(\gamma_\pm)$ are
the arbitrary isotropic particle distributions $\int f_\pm
d\gamma_\pm=1$, $n_\pm$ and $\gamma_\pm$ are the $e^\pm$ number
densities and Lorentz factors. An analytical expression for the
angle-averaged emissivity per pair of particles,
   \begin{equation}
H(\nu,\gamma_+,\gamma_-)=\int d\cos\theta^*\,{\gamma^{*2}\beta^*\over
\gamma_+\gamma_-}{d\sigma\over d\nu}(\nu,\gamma_+,\gamma_-,\cos\theta^*),
                                                            \label{7.3}
   \end{equation}
was obtained by Svensson (1982), here ${d\sigma\over d\nu}$ is the
differential cross section for emission of a photon with LS energy
$\nu$.

\section {Calculations and Analysis}
\begin{figure} 
   \vskip -4cm
   {          
      \psfig{file=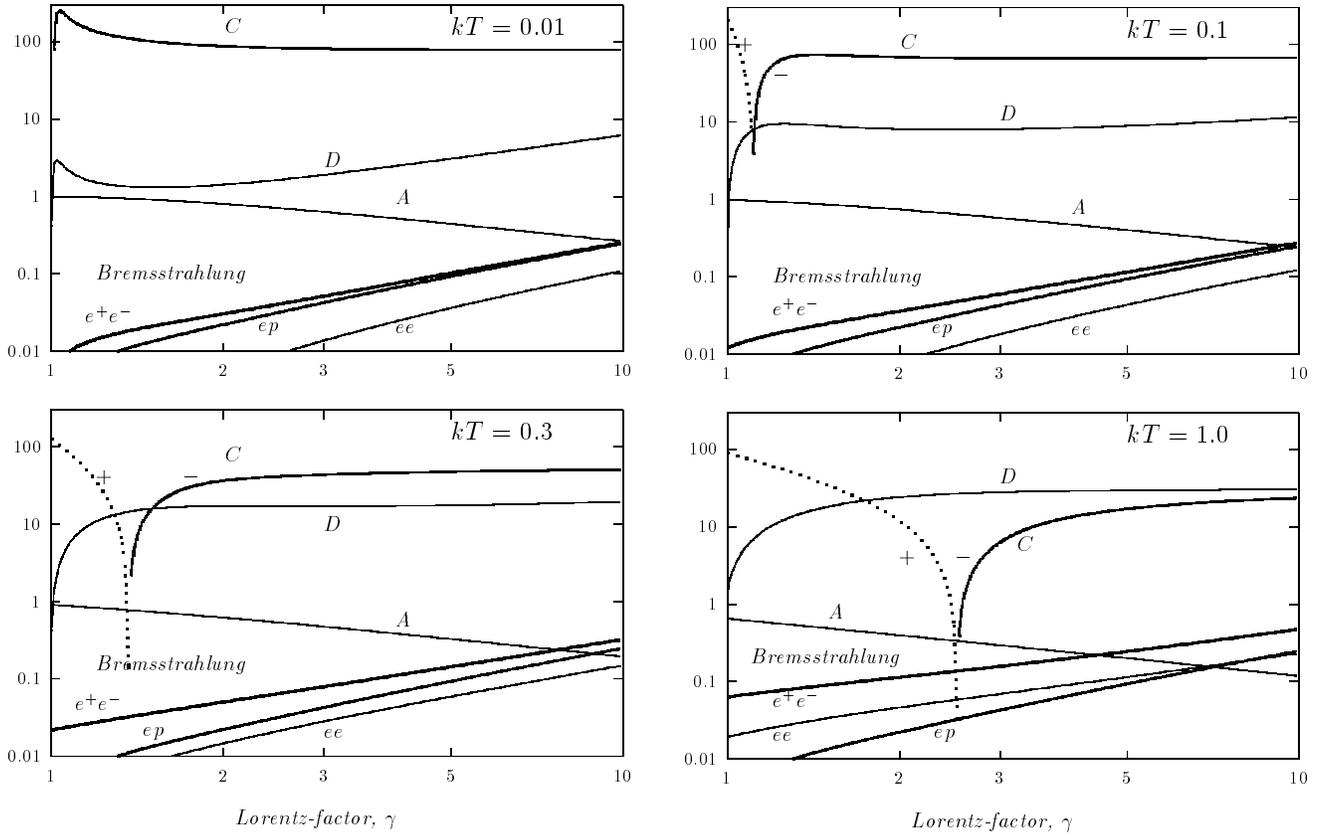,width=21cm,height=29.7cm,clip=}
   }

   \vskip -14.5cm
   \caption{\label{f1}
Shown are the calculated annihilation rate (A), energy losses due to
bremsstrahlung ($e^+e^-$, $ee$, and $ep$) as well as Coulomb energy
losses (C) and dispersion coefficients (D) in thermal hydrogen
plasmas. All values are provided dimensionless, in units $n_e\pi
r_e^2$. Low energy particles gain energy in Coulomb scattering with
plasma particles that appears as a sign change and shown by bold dotted
lines.
} 
   \end{figure}

\begin{figure} 
   \vskip -4cm
   {          
      \psfig{file=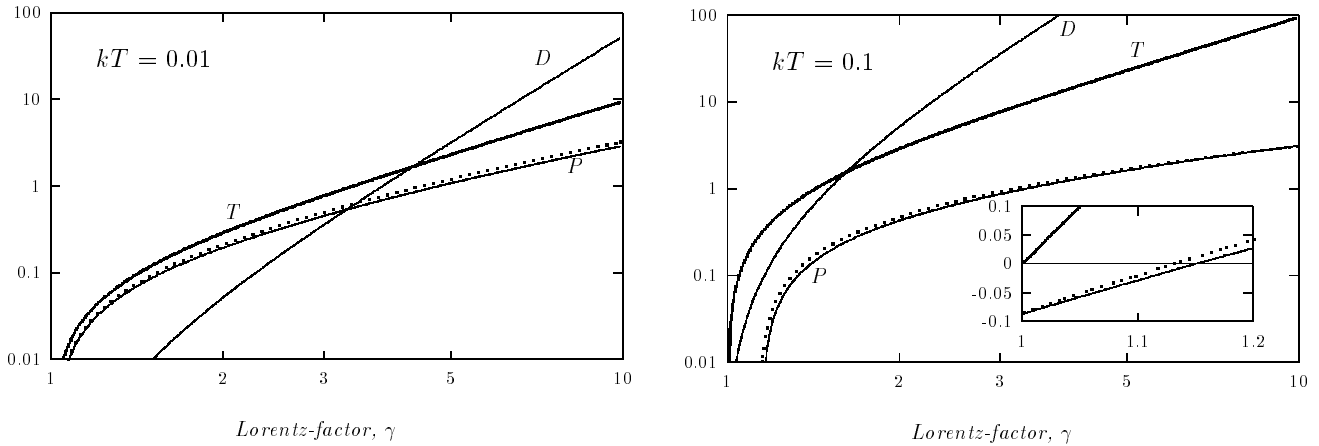,width=21cm,height=29.7cm,clip=}
   }
   \vskip -19.8cm
   \caption{\label{f2}
Energy losses due to the Comptonization. The losses are shown in the
units $n_e\pi r_e^2$, the photon number density has been taken equal to
that of the plasma electrons $n_\gamma=n_e$. The thin lines (P) show the
energy losses on Planck's photons. The delta-function approximation of
Planck's distribution with $\omega_0=2.7kT$ is shown by the dotted
lines. Thick solid lines (T) show the Thomson limit of the Compton
scattering. The dispersion coefficient in the Thomson limit is shown by
the solid line (D). The inset shows the enlarged low-energy part
without the dispersion.
} 
   \end{figure}

\begin{figure} 
   \vskip -4cm
   {          
      \psfig{file=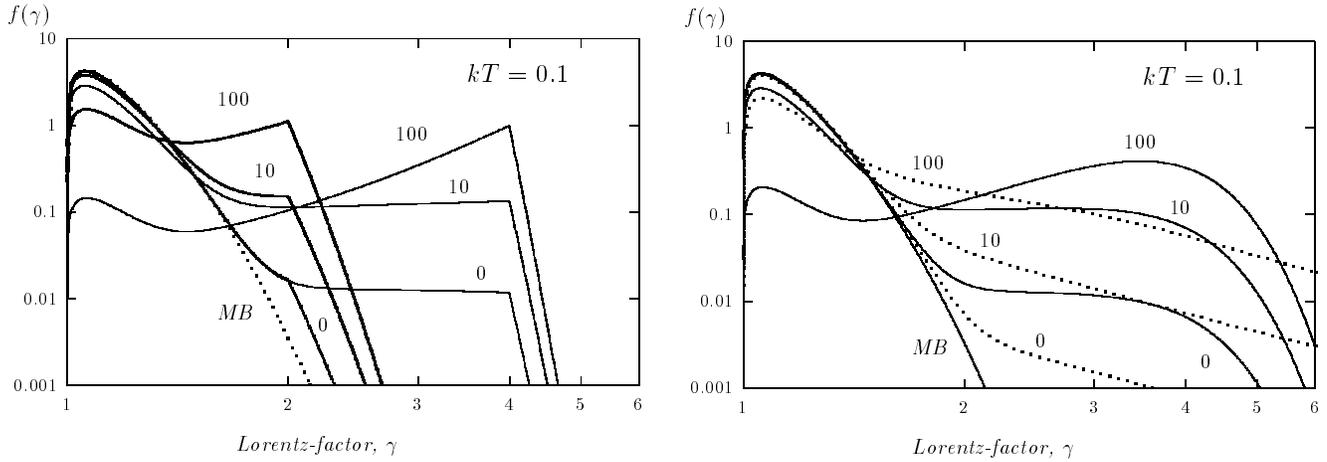,width=21cm,height=29.7cm,clip=}
   }
   \vskip -19.8cm
   \caption{\label{f3}
The distorted positron distribution $f(\gamma)$ for an
electron temperature $kT=0.1$ with and without positron escape
$E(\gamma)=0$, 10, 100. The left panel shows the positron distributions
for the case of monoenergetic source function $\tilde
S(\gamma)=\delta(\gamma-\gamma_0)$ with $\gamma_0$=2 (thick lines) and
4 (thin lines). A Maxwell-Boltzmann distribution is shown by a dotted
line. The right panel shows the positron distributions for the cases of
power-law $\tilde S(\gamma)=2/\gamma^3$ (dotted lines) and Gaussian
$\tilde S(\gamma)=\exp[-(\gamma-4)^2]/ \sqrt{\pi}$ (solid lines). A
Maxwell-Boltzmann distribution (MB) is also shown.
} 
   \end{figure}

\begin{figure} 
   \vskip -4cm
   {          
      \psfig{file=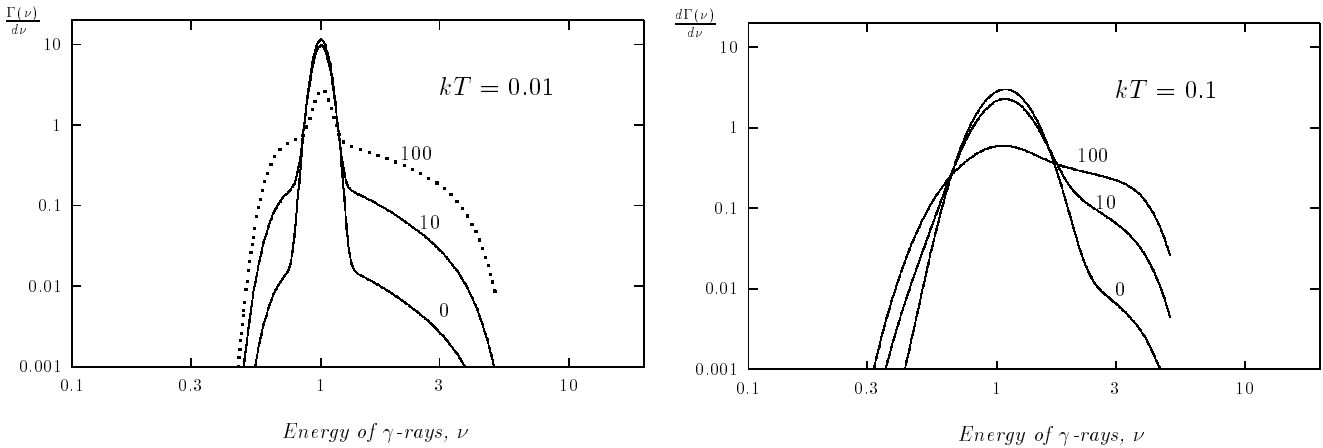,width=21cm,height=29.7cm,clip=}
   }
   \vskip -19.8cm
   \caption{\label{f4}
The spectra of photons from annihilation of positrons
with Maxwellian electrons for $kT=0.01$ and 0.1 with and without
positron escape $E(\gamma)=0$, 10, 100. The source function of
positrons was taken a Gaussian $\sim\exp[-(\gamma-4)^2]$. The spectra
are provided dimensionless, in units $n_-n_+\pi r_e^2$, where $n_\pm$
is the $e^\pm$ number density and $r_e$ is the classical electron
radius.
} 
   \end{figure}

\begin{figure} 
   \vskip -4cm
   {          
      \psfig{file=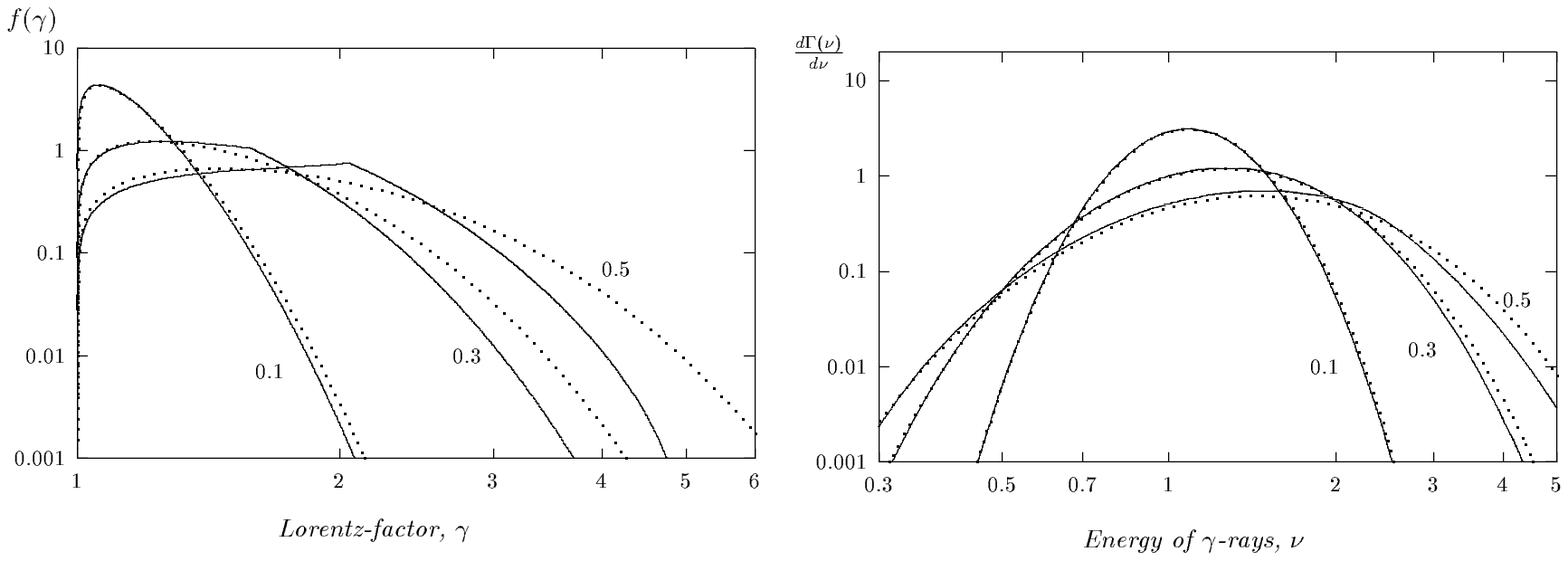,width=21cm,height=29.7cm,clip=}
   }
   \vskip -19.8cm
   \caption{\label{f5}
The distorted positron distribution $f(\gamma)$ for values of electron
temperature $kT=0.1$, 0.3, and 0.5 (left panel). The positron escape
rate was taken the same $E=50\beta$ for all three cases.
Corresponding Maxwell-Boltzmann distributions are shown by dotted lines.
The right panel shows the spectra of photons from annihilation of positrons
with Maxwellian electrons for $kT=0.1$, 0.3, and 0.5.
The line shapes for annihilation of Maxwellian positrons with Maxwellian
electrons are shown by dotted lines.
} 
   \end{figure}

The rates obtained in the paper were integrated over the Maxwellian
distribution in order to compare with well-known results for the thermal
plasma. The annihilation rate was tested with annihilation rate of an
$e^+e^-$ plasma (Ramaty \& M\'esz\'aros 1981), bremsstrahlung energy
losses were compared with $e^+e^-$-, $ee$-, and $ep$-bremsstrahlung
luminosities of thermal plasmas (Haug 1985c). Two more tests on Coulomb
energy losses and bremsstrahlung were carried out with calculations by
Dermer \& Liang (1989). An excellent agreement was found. Compton energy
loss eq.~(\ref{6.6})--(\ref{6.7}) coincides with the Thomson limit as
$\omega\to0$. Besides, we have found that the formulas obtained can be
also successfully applied for the calculation of the bremsstrahlung
luminosity and annihilation rate of the thermal plasma by replacing the
positron Lorentz factor with the average one over the Maxwellian
distribution $\langle\gamma\rangle_{kT}=3kT+[K_1(1/kT)/K_2(1/kT)]$.

The relevant energy loss rates ($-d\gamma/dt$) and annihilation rate per
one positron are shown in Fig.~\ref{f1} and \ref{f2}. All values are
provided dimensionless, in units $n_e\pi r_e^2$, the Coulomb logarithm
was taken a constant $\ln\Lambda=20$. M\o ller and Bhabha energy losses
show negligible difference and dominate over the others except Compton
scattering, which is quite effective and can prevail at large Lorentz
factors of positrons (electrons). Low energy particles gain energy in
Coulomb scattering with thermal electrons that appears as the sign
change of $d\gamma/dt$. Energy losses due to bremsstrahlung are
negligible in comparison with others. Annihilation rate is small in
comparison with the relaxation rate, so that most of positrons
annihilate after their distribution approaches the steady-state one.

The energy losses due to Compton scattering (Fig.~\ref{f2}) have been
calculated in the Thomson limit eq.~(\ref{6.8}) and in the
Klein-Nishina regime for a Planckian spectrum
$f_\gamma(\omega)=(e^{\omega/kT}-1)^{-1}/2.404(kT)^3$, and the
$\delta$-function approximation. The energy loss rates due to the
Comptonization on Planck's photons are shown for two photon
temperatures, the $\delta$-function approximation of Planck's
distribution with $\omega_0=2.7kT$ gives similar results. For the clear
comparison with Fig.~\ref{f1} the photon number density have been taken
equal to that of the plasma electrons $n_\gamma=n_e$ easily
generalizing for an arbitrary $n_\gamma$ by trivial vertical shift of
the curves. For the coherence, in all calculations the energy density of
photons $U_{ph}$ was taken equal to that of Planck's distribution
$\approx 2.7kTn_\gamma$. Shown also is the dispersion coefficient
calculated in the Thomson limit eq.~(\ref{6.9}). The radiation can
provide some heating for the cold particles similar to that in the
Coulomb scattering. Very low-energy particles gain energy due to
Comptonization that appears as a sign change of the energy losses (see
the inset in Fig.~\ref{f2}). Clearly, the effect results from using the
Klein-Nishina cross section.

At small positron Lorentz factors, the Coulomb energy losses dominate
the losses due to Comptonization over the variety of photon temperatures
and densities (cf.\ Fig.~\ref{f1} and \ref{f2}). Qualitatively it means
that high photon density leads to the cooling of plasma preferentially
through high-energy particles. Herewith, the Coulomb scattering mixes
particles so that the plasma remains nearly Maxwellian. Therefore the
energy losses due to Comptonization would be only important for the
high-energy tail of the particle distribution, which becomes narrower.
The precise shape of the distribution would be driven by the balance of
income and outcome energy fluxes.

Positrons could be injected into the hydrogen plasma volume by an
external source or produced in the bulk of the plasma. In the latter
case the form of the source function is governed by the nature of the
processes involved. Electron-positron pair production in
$ep$-collisions becomes possible when the electron interacting with a
stationary proton has the Lorentz factor exceeding 3, for
$ee$-collisions one should exceed 7 when one interacting particle is at
rest. If the pair is to be produced in two-photon collisions, the
photon energies, $\omega_i$, and the relative angle, $\theta$, must
satisfy the condition $\omega_1\omega_2>2/(1-\cos\theta)$. Low plasma
temperature is consistent with a small positron fraction in the plasma
since the positrons could be produced by the relatively small number of
head on collisions of energetic photons  and/or electrons from the tail
of Maxwellian distribution.

If the particle production is not balanced by annihilation it could
lead to escape of $e^+e^-$-plasma, since the gravitation near a compact
object can't prevent pairs from escaping. Two independent mechanisms,
at least, diffusion and the radiation pressure result in escaping of
particles from the plasma volume. We, therefore, explore these factors
separately. If particles escape due to the radiation pressure, it is
natural to suppose that the escape probability $E(\gamma)$ is a weak
function of the particle Lorentz factor, we thus put it a constant. In
the case the escape is of diffusive origin, the diffusion coefficient
is a function of particle speed ${\mathcal D}\sim\beta$. We thus
consider two functional forms for the escape probability $E\sim\beta$,
and $E\sim\sqrt{\beta}$ which simulates the case when both
mechanisms operate simultaneously.

Calculations of the distorted function $f(\gamma)$ have been made
(Fig.~\ref{f3}) for the source function in the form of monoenergetic
distribution $\tilde S(\gamma)=\delta(\gamma-\gamma_0)$, power-law
$\tilde S(\gamma)=2/\gamma^3$, and Gaussian $\tilde
S(\gamma)=\exp[-(\gamma-4)^2]/ \sqrt{\pi}$. The escape rate was taken
energy-independent $E=0$, 10, and 100 in units $n_e\pi r_e^2$, which is
negligible, medium and very high in comparison with the time scale of
the Coulomb energy losses (cf.\ Fig.~\ref{f1}). It demonstrates an
effect of blowing away of (unbound) electron-positron pairs by
radiation pressure.

The behavior of the solution $f(\gamma)$ depending on the injection
function and escape rate is quite clear from the Figure. One can
show that the right side of the eq.~(\ref{2.8}) and (\ref{2.3}) is
negligible at $\gamma\to1$, the solution is therefore Maxwellian-like.
Beginning from some point, the term $\int_1^\gamma (A+E)f \,d\gamma'$
becomes non-negligible that leads to some increasing of the derivative
$g'(\gamma)$ and deviation of the solution from Maxwellian. Thus, a
bump is forming. At some Lorentz factor the last term in the right side
of eq.~(\ref{2.8}) and (\ref{2.3}) is switching on, which leads to some
decreasing of the derivative or could even change it to a negative
value. At large Lorentz factors the right side of the equations again
approaches zero (see eq.~[\ref{2.4}]). Generally, if the energy of
injected particles essentially exceeds the average one of plasma
particles it leads to an extended tail, while the correct normalization
of the whole solution thus requires some deficit at low energies.

Typical spectra of photons from annihilation of these positrons with
Maxwellian electrons are shown in Fig.~\ref{f4} for electron
temperatures $kT=0.01$, and 0.1. It is seen that as plasma temperature
grows the annihilation line widens, its height decreases and
distortions of its shape become relatively more intensive.

Another case is shown in Fig.~\ref{f5}. The distorted functions were
calculated for electron temperatures $kT=0.1$, $0.3$, and $0.5$ while
the escape probability in all cases was taken the same $E=50\beta$ (in
units $n_e\pi r_e^2$). The actual values of the escape rate in these
cases could be inferred from the value of the integral
$\int_1^\infty (A+E)f\,d\gamma'$, which is equal to $\approx24$,
$\approx35$, and $\approx41$, correspondingly. Particle injection was
taken monoenergetic with energy equal to the average energy of plasma
electrons. In all cases, the escape leads to some deficit of energetic
particles in the tail of distribution, while the particle injection
appears as a bump. Although the distributions of positrons differ from
Maxwellians, their annihilation with thermal electrons does not lead to
large distortions of the annihilation line form. This latter is very
similar to the line from annihilation of two Maxwellian distributions.

Although only few cases have been discussed, the performed calculations
have shown that the functional dependence of the escape rate is not
very important. In all three cases $E=const$, $\sim\beta$, and
$\sim\sqrt{\beta}$ we obtained similar results for the same injection
function, the difference appears only at very low temperatures
$kT\la0.05$. It is quite clear, since $\beta$ increases
from 0 to $\approx1$ in a narrow region $\gamma=1-1.4$ remaining
further a constant. The particle distribution actually depends on
the value $\int_1^\infty (A+E)f\,d\gamma$, energy of the injected
particles and their distribution (cf.\ Figures \ref{f3} and
\ref{f5}).  In absence of the particle injection, the escape of
particles operates as an additional mechanism for the plasma cooling.

\section {Nova Muscae and 1E 1740.7--2942}
Recent observations with {\it SIGMA} telescope have revealed
annihilation features in the vicinity of $\sim500$ keV in spectra of
two Galactic black hole candidates, 1E~1740.7--2942 (hereafter the
1E source; Bouchet et al.\ 1991; Sunyaev et al.\ 1991; Churazov et
al.\ 1993; Cordier et al.\ 1993), and Nova Muscae (Sunyaev et
al.\ 1992; Goldwurm et al.\ 1992). During all periods of observation
the hard X-ray emission, 35--300 keV, was found to be consistent
with the same law. Observations of Nova Muscae after the X-ray flare
(January 9, 1991) are well fitted by a power law of index $2.4-2.5$ or
by Sunyaev-Titarchuk (1980) model with $kT\approx55-75$ keV and
$\tau\approx0.4-0.5$ in the disc geometry, the spectrum of the 1E
source is well described by Sunyaev-Titarchuk model with
$kT\approx35-60$ keV and $\tau\approx1.1-1.9$. Meanwhile soft
$\gamma$-ray emission of these sources seems to be highly variable.

\begin{figure}[tbh]
   {      
      \psfig{file=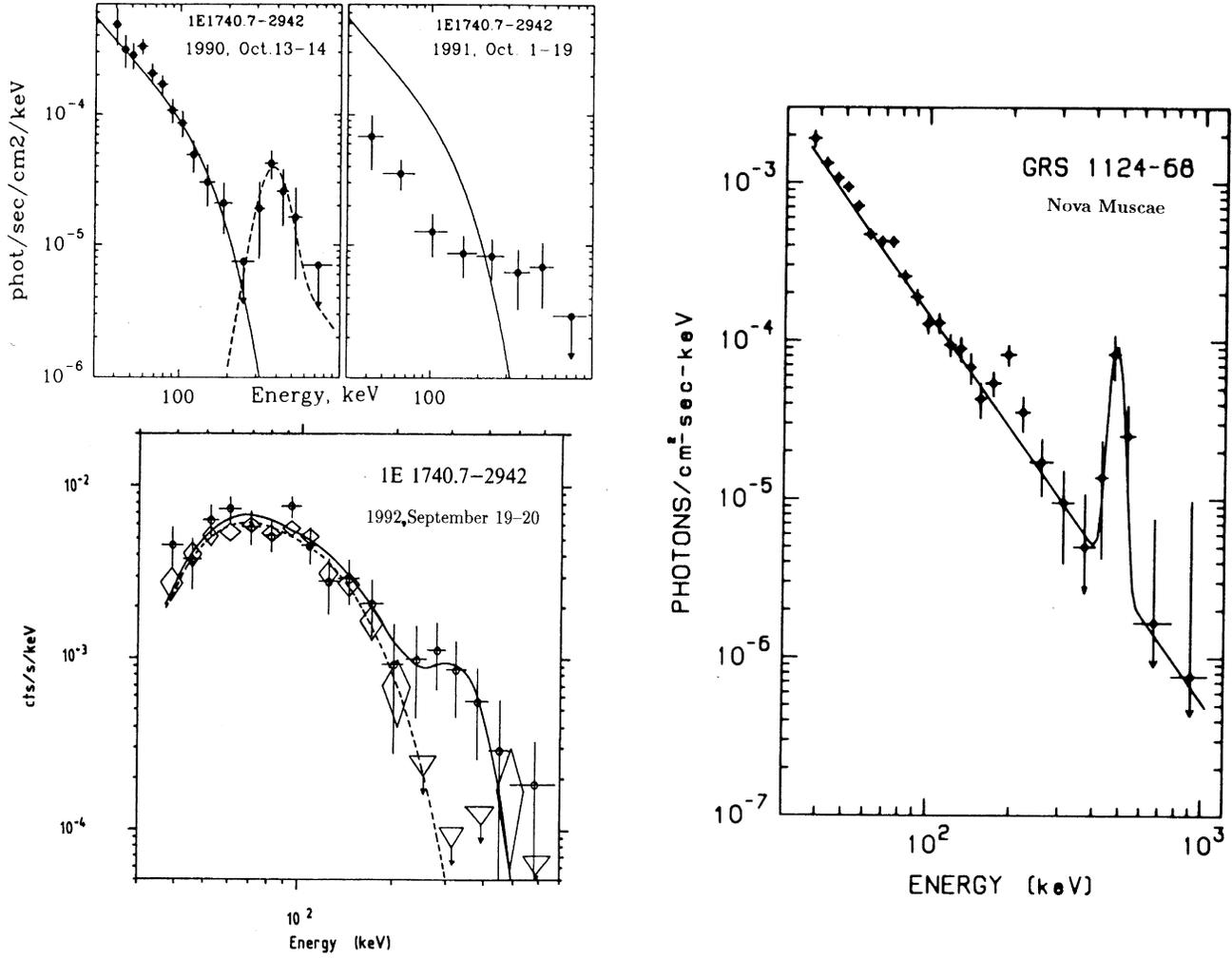,width=18cm,height=14cm,clip=}
   }
   \caption{\label{f6}
Energy spectra of the 1E source (Bouchet et al.\ 1991; Churazov et
al.\ 1993, 1994; Cordier et al.\ 1993) and Nova Muscae (Goldwurm et
al.\ 1992) observed by {\it SIGMA} are shown together with fits of the
authors. For September 1992 flare shown is counts s$^{-1}$ keV$^{-1}$.
The dashed line in the upper left panel shows the annihilation
line shape for Gaussian-like injection, $\sim\exp[-(\gamma-4)^2]$, of
energetic particles into the thermal plasma of $kT=35$ keV for $E/A=20$.
The line is shifted left to approach the data.
}
    \end{figure}

During the last 13 hr of a 21 hr observation on January 20--21, 1991, a
clear emission feature around 500 keV was found in the spectrum of Nova
Muscae (Fig.~\ref{f6}), with a line flux of $\approx6\times10^{-3}$ photons
cm$^{-2}$ s$^{-1}$, and an intrinsic line width $\la58$ keV (Sunyaev et
al.\ 1992; Goldwurm et al.\ 1992). Since the first 8 hr of the
observation did not give a positive detection, the inferred rise time
is equal to several hours. The next observation, held on February 1--2,
did not show this feature restricting the lifetime to $\la10$ days.

The Galactic center region was intensively monitored by {\it SIGMA}
telescope since 1990. Three times during these years a broad excess was
observed in the 200--500 keV region (Fig.~\ref{f6}).

In an observation performed between 1990 October 13 and 14, a
spectacular unexpected feature was found in the 1E emission spectrum,
the corresponding flux was estimated at $(0.9-1.3)\times10^{-2}$
photons cm$^{-2}$ s$^{-1}$, with a line width of $180-240$ keV (Bouchet
et al.\ 1991; Sunyaev et al.\ 1991). The observations of this region
performed two days before (on October 10--11), and a few hours after
(October 14--15) did not exhibit any spectral feature beyond 200 keV.
The total duration of this state is estimated between 18 and about 70
hr.

Seven October 1991 observations have shown an evident excess at high
energies, while the source was in a low state (Churazov et al.\ 1993).
The excess was observed during 19 days and was not so intensive as in
October 1990, the average flux was $(1.9\pm0.6)\times10^{-3}$ photons
cm$^{-2}$ s$^{-1}$ in the 300--600 keV region.

The 1992 September 19--20 observational session (Sep.\ 19.42--20.58)
showed a feature beyond 200 keV (Cordier et al.\ 1993), which
resembles  that of 1990 October 13--14. The line flux was estimated as
$4.28_{-1.50}^{+2.70}\times10^{-3}$ photons cm$^{-2}$ s$^{-1}$. The
previous (Sep.\ 18.59--19.30) and the next (Sep.\ 22.57--23.14)
sessions did not show any evidence for emission in excess of
200 keV, restricting the lifetime of the state between 27 and about 75
hr, while the rise time approaches probably several hours.

The spectral features observed by {\it SIGMA} are, commonly believed,
related to electron-positron annihilation. Relatively small line widths
imply that the temperature of the emitting region is quite low,
$kT\approx35-45$ keV for 1E and 4--5 keV for Nova Muscae. Since
the hard X-ray spectra $<300$ keV showed no changes, most probably
that electron-positron pairs produced somewhere close to the central
object were injected into surrounding space where they cool and
annihilate.  Radiation pressure of a near-Eddington source alone can
accelerate $e^+e^-$-plasma up to the bulk Lorentz factor of
$\gamma_0\sim2-5$ (Kovner 1984), while Comptonization by the emergent
radiation field (Levich \& Syunyaev 1971) could provide a mechanism for
cooling the pairs which further annihilate ``in flight" (for a
discussion see also Gilfanov et al.\ [1991, 1994]). If there is enough
matter around a source, then particles slow down due to Coulomb energy
losses and annihilate in the medium. We explore further this last
possibility  by checking whether the inferred parameters of the
emitting region are consistent with those obtained by other ways.  
We assume single and short particle ejection on a timescale of hours.
It seems reasonable: since the ejection would probably impact on the
whole spectrum, longer spectral changes would be observable.

Suggesting that the energetic particles slow down due to Coulomb
scattering in the surrounding matter, one can estimate its (electron) number
density
   \begin{equation} 
n_-\approx{\gamma_0-1\over\pi r_e^2 c\,\Delta_i}\left({d\gamma\over
dt}\right)^{-1},                                            \label{8.1}
   \end{equation}
where $\gamma_0$ is the initial Lorentz factor of the plasma stream,
$c$ is the light speed, and $\Delta_i$ is the characteristic time scale
of the annihilation line appearance. The Coulomb energy loss rate in a
medium of $kT\le0.1$ is $(d\gamma/d t)\approx70-100$ (see
Fig.~\ref{f1}). Taking a reasonable value for the bulk Lorentz factor,
$\gamma_0\approx3$ (e.g., Kovner 1984), one can obtain estimations of
the order of magnitude as
$n_-\approx2.2\times10^7$~cm$^{-3}$~$(\Delta_i/2$~days$)^{-1}$ for the 1E
source, and
$n_-\approx1.5\times10^8$~cm$^{-3}$~$(\Delta_i/5$~hr$)^{-1}$ for Nova
Muscae.

If the energetic particles were injected into the medium only once,
then the annihilation feature lifetime $\Delta_d$ is directly connected
with annihilation rate as $\Delta_d^{-1}\approx\pi r_e^2 c\,n_-A(\gamma)$. 
It yields one more estimation of the number density in the emitting region
   \begin{equation}
n_-\approx
{1\over\pi r_e^2c\,\Delta_d\,A(\gamma)}\approx
1.55\times10^9{\rm\ cm}^{-3}\ \left({\Delta_d\over 1{\rm\ day}}\right)^{-1}.
                                                            \label{8.2}
   \end{equation}
Annihilation rate $A(\gamma)$ is a weak function of $\gamma$ (see
Fig.~\ref{f1}) and we can take it equal to a constant $A=A(1)\approx1$.
Total duration of the hard state is $\Delta_d\approx18-70$ hr for the
1E source and $\Delta_d\le10$ days for Nova Muscae, that gives
$n_-\approx(5-20)\times10^8$ cm$^{-3}$ and
$n_-\approx1.5\times10^8$~cm$^{-3}$~$(\Delta_d/10$~days$)^{-1}$,
correspondingly. The values obtained from eq.~(\ref{8.1})--(\ref{8.2})
restrict the electron number density in the volume where particles slow
down and annihilate.

Being equated eq.~(\ref{8.1})--(\ref{8.2}) give an obvious relation
between the time scales
   \begin{equation} 
{\Delta_d\over\Delta_i}={1\over A(\gamma_0-1)}{d\gamma\over dt}.\label{8.3}
   \end{equation}
Therefore, to be consistent with the annihilation lifetime the
annihilation rise time for the 1E source should be $\Delta_i\approx1-2$
hr. This is supported by the 1992 September 19--20 observation when the
annihilation rise time was restricted by a few hours.

\begin{table}[t]
   \caption{Observational data and parameters of the emitting region.}
      \begin{tabular}{llll}
         \noalign{\smallskip}
         \hline
         \noalign{\smallskip}

& \multicolumn{2}{c}{1E 1740.7--2942} & \\
         \noalign{\smallskip}
         \cline{2-3}
         \noalign{\smallskip}
& 1990 Oct.\ 13--14 & 1992 Sep.\ 19--20 &
                     \raisebox{1.5ex}[0pt]{\ Nova Muscae} \\
         \noalign{\smallskip}
         \hline
         \noalign{\smallskip}

Annihilation rise time, $\Delta_i$
  & $\la2$ days (1--2 hr)$^\ast$ & few hours & $\sim5$ hr \\

Annihilation lifetime, $\Delta_d$
  & 18--70 hr        & 27--75 hr       & $\la10$ days\\

Annihilation photon flux, $F_{500}$ (photons cm$^{-2}$ s$^{-1}$)
  & $10^{-2}$     & $4.3\times10^{-3}$ & $6\times10^{-3}$\\

Total line flux, $L_{500}$ (photons s$^{-1}$)
  & $8.6\times10^{43}$ (at 8.5 kpc) & $3.7\times10^{43}$
                                       & $7.2\times10^{41}$ (at 1 kpc)\\

Line width, $W$ (keV) & 240   & 180    & 40 \\

Plasma temperature, $kT_e$ (keV)
   & \multicolumn{2}{c}{$35-45$}       & $3-4$ \\

Column density, $N_H$ (cm$^{-2}$)
   & \multicolumn{2}{c}{$\sim10^{23}$} & $\sim10^{21}$  \\

Coulomb energy loss rate, $d\gamma/dt$
   & \multicolumn{2}{c}{70}            & 100   \\

Annihilation rate, $A$
   & \multicolumn{2}{c}{1}             & 1     \\

Electron number density, $n_-$ (cm$^{-3}$)
 &\multicolumn{2}{c}{$(5-20)\times10^8$} & $1.5\times10^8$ \\

Size of the emitting region, $\lambda$ (cm)
 & \multicolumn{2}{c}{$(1.1-20)\times10^{13}$} & $(1.3-50)\times10^{13}$\\
         \noalign{\smallskip}
         \hline
      \end{tabular}
\begin{list}{}{}
\item[$^\ast$] Our estimation.
\end{list}
\end{table}

The size of the emitting region $\lambda$ could be estimated from a
simple relation $n_+\lambda^3\sim\Delta_d L_{500}/2$ if we assume the
upper limit for the positron number density $n_+\le n_-$. It gives
$\lambda\ga1.34\times10^{13}$~cm~$(\Delta_d/1$~day$)^{2/3}
\approx(1.1-2.7)\times10^{13}$ cm for 1E and
$\lambda\ga1.3\times10^{13}$~cm~$(\Delta_d/10$~days$)^{2/3}$ for Nova
Muscae\footnote{We took $n_+\le
n_-\approx1.5\times10^8$~cm$^{-3}$~$(\Delta_d/10$~days$)^{-1}$.}, which
are well inside of the upper limits $\lambda<
c\Delta_i\approx2.2\times10^{14}$~cm~$(\Delta_i/2$~hr$)$ and
$\le5\times10^{14}$ cm, correspondingly.
From the above consideration follows that emitting regions
in both sources are optically thin and do not affect the Comptonized
spectra at $<300$ keV nor the annihilation line form. Experimental
data and the estimated parameters are summarized in Table 1.

The column density of the medium where injected particles slow down and
annihilate should exceed the value $N_H\sim\lambda n_-$, which follows
from previous estimations for $\lambda$ and $n_-$, viz.\
$2.1\times10^{22}$~cm$^{-2}$~$(\Delta_d/ 1$~day$)^{-1/3}\la N_H<
c\Delta_in_-\approx1.1\times10^{23}$~cm$^{-2}$ for 1E, where we took
into account eq.~(\ref{8.3}), and
$N_H\ga2\times10^{21}$~cm$^{-2}$~$(\Delta_d/ 10$~days$)^{-1/3}$ for
Nova Muscae. The total column density of the gas cloud measured along
the line of sight, where the 1E source embedded, is high enough
$N_H\approx3\times10^{23}$ cm$^{-2}$ (Bally \& Leventhal 1991; Mirabel
et al.\ 1991). Note that recent {\it ASCA} measurements of the column
density {\it to this source} give a best fit value
$N_H\approx8\times10^{22}$ cm$^{-2}$ (Sheth et al.\ 1996). For Nova
Muscae the corresponding value is $N_H\sim10^{21}$ cm$^{-2}$ (Greiner
et al.\ 1991), less or marginally close to the obtained lower
limit. If, on contrary,
one suggests $n_+\ll n_-$ it yields a condition
$N_H\gg2\times10^{21}$~cm$^{-2}$~$(\Delta_d/ 10$~days$)^{-1/3}$, which
considerably exceeds the measured value.

These estimations put us on to an idea that the 500 keV emission
observed from Nova Muscae was coming from $e^+e^-$-plasma jet
($n_+\approx n_-$) rather than from particles injected in a gas 
cloud\footnote{A possibility that Nova Muscae lies in front of a large
gas cloud can not be totally excluded. In this case, particles could be
injected into this cloud, away from the observer.}
($n_+\ll n_-$), therefore, particles have to annihilate ``in flight"
producing a relatively narrow line blue- or red-shifted dependently on
the jet orientation. If so, then our estimation of the electron number
density $n_-$ from annihilation time scale is related to the average
electron/positron number density in the jet, its total volume is of
$\lambda^3\sim2\times10^{39}$~cm$^3$~$(\Delta_d/10$~days$)^2$. The
reported 6\%--7\% redshift of the line centroid (Goldwurm et
al.\ 1992; Sunyaev et al.\ 1992) supports probably the
annihilation-in-jet hypothesis, although authors noted that statistical
significance of this shift is not very high. The large size of the
emitting region and a small width of the line, both except the
gravitational origin of the redshift, since in this case the
annihilation region have to be quite close to the central object
$\sim10 R_g$ where typical flow velocities should result in a much
broader line (Gilfanov et al.\ 1991). The Compton scattering of the
anisotropic emergent radiation could provide effective mechanism for
blowing away and acceleration of $e^+e^-$-pair plasma (e.g.,
Kovner 1984; Misra \& Melia 1993) cooling it at the same time. Since
the maximal energy during the X-ray flare of Nova Muscae released near
$\sim1$~keV (Greiner et al.\ 1991), the average kinetic energy per
particle should be nearly the same (which is consistent with 
the small line width).

The case of the 1E source is not definitively clear, because our
estimations give $n_-\ga n_+$ in the emitting region. Two flares,
October 1990 and September 1992, have shown very similar time scales,
spectra and photon fluxes, which are consistent with single injection
of energetic particles into the thermal (hydrogen) plasma. Meanwhile,
the redshift of the line $\sim25$\% reported by authors (Bouchet et
al.\ 1991; Sunyaev et al.\ 1991; Cordier et al.\ 1993) implies that
positrons probably annihilate in a plasma stream moving away from the
observer. The estimation of the size of the emitting region ruled
out its gravitational nature, since it is too large in comparison with
gravitational radius of a stellar mass black hole. A natural
explanation of this controversial picture is that the propagating
plasma stream captures matter from the source environment and
annihilation occurs in a moving plasma volume. In this case the
estimation of the electron number density $n_-$ is related to the
average electron number density in the jet,
$\lambda^3\ga2.4\times10^{39}$~cm$^3$~$(\Delta_d/1$~day$)^2$ gives its
total volume, and the jet length has to be of the order of
$\sim0.2c\Delta_d\approx5.2\times10^{14}$~cm~$(\Delta_d/1$~day$)$.

While a part of the $e^+e^-$-pair probably annihilate in a thermal plasma
near the 1E source producing the broad line, the remainder could escape
into a molecular cloud, which was found to be associated with the
1E source (Bally \& Leventhal 1991; Mirabel et al.\ 1991). The time
scale for slowing 
down\footnote{
The corresponding annihilation lifetime $\Delta_d$
(eq.~[\ref{8.2}]--[\ref{8.3}]) was obtained for thermal plasma and is
not valid for the cold medium where positrons mostly annihilate in the 
bound (positronium) state.
}
due to the scattering could be obtained from eq.~(\ref{8.1}). Taking
$\sim10^5$ cm$^{-3}$ for the average number density of the molecular
cloud near 1E (Bally \& Leventhal 1991; Mirabel et al.\ 1991) one
gets $\Delta_i\la1$ year, the same as that obtained by Ramaty et
al.\ (1992). The size of the turbulent region in the cloud caused by
propagation of a dense jet should be of the same order. It agrees well
with the length 2--4 ly (15--30 arcsec at the 8.5 kpc distance) of a
double-sided radio jet from the 1E source found recently with the VLA
(Mirabel et al.\ 1992).

If the lines from the 1E source (Fig.~\ref{f6}) were produced by
continuous injection of energetic particles, then the observations of
the narrow 511 keV line emission from the Galactic center allows to put
an upper limit on the particle escape rate into the interstellar
medium. Recent reanalysis of {\it HEAO 3} data has shown that under
suggestion of a single point source at the Galactic center narrow
line intensities are $F_{511}=(1.25\pm0.18)\times10^{-3}$ photons
cm$^{-2}$ s$^{-1}$ for the fall of 1979 and
$F_{511}=(0.99\pm0.18)\times10^{-3}$ photons cm$^{-2}$ s$^{-1}$ for the
spring of 1980 (Mahoney et al.\ 1994). Taking $\tau_0=1$ yr for the
positron lifetime in $10^5$ cm$^{-3}$ dense cold molecular cloud
(Ramaty et al.\ 1992), and suggesting  one hard state of $\Delta_d\ga2$
days long per period $\tau_0$, one can obtain an escape rate
$E/A\approx{F_{511}\,\tau_0\over F_{500}\,\Delta_d}\la20$, where we took
$F_{500}=10^{-2}$ photons cm$^{-2}$ s$^{-1}$ (see Table 1). This is
consistent with the upper limits of 1990 October 13--14 spectrum and
the two most energetic points in 1992 September 19--20 spectrum. The
dashed line in 1990 October 13--14 spectrum (Fig.~\ref{f6}) shows the
annihilation line shape for Gaussian-like injection,
$\sim\exp[-(\gamma-4)^2]$, of energetic particles into the thermal
plasma of $kT=35$ keV for $E/A=20$. The longest hard state ($\sim19$
days) with the average flux of $F_{500}\approx2\times10^{-3}$ photons
cm$^{-2}$ s$^{-1}$ observed in October 1991 places the upper limit at
almost the same level $E/A\approx10$.

\section {Conclusion}
We have presented the accurate formulas in the form of a simple expression
or an one-fold integral for the energy losses and gains of particles
scattered by a Maxwell-Boltzmann plasma. The processes concerned are the
Coulomb scattering, $e^+e^-$-, $ee$- and $ep$-bremsstrahlung as
well as Comptonization in the Klein-Nishina regime.

The problem of positron propagation is treated in a Fokker-Planck
approach, which can be easily generalized to include inelastic
processes, stochastic acceleration etc. We have shown that the escape
of positrons in the form of pair plasma has an effect on the positron
distribution causing, in some cases, a strong deviation from a
Maxwellian. When the energy of injected particles essentially exceeds
the average one of plasma particles, the deviation appears as a deficit
at lower energies and an extended tail of the distribution that leads
to a widening and smoothing of the annihilation feature in the spectrum.
In the case where the energy of particles
injected is close to the average energy of plasma particles, the
deviation appears as an injection bump and a deficit in the tail of the
distribution. Meanwhile, it does not lead to visible distortions of
the annihilation line shape which is similar to that of thermal
plasmas.

The performed calculations allow us to obtain reliable
estimations of the electron number density and the size of the emitting
regions in Nova Muscae and the 1E~1740.7--2942 source, suggesting that
spectral features in 300--600 keV region observed by {\it SIGMA}
telescope are due to the electron-positron annihilation in thermal
plasma. We conclude that in the case of Nova Muscae the observed
radiation is coming from a pair plasma jet, $n_+\approx n_-$, rather
than from a gas cloud. The case of 1E~1740.7--2942 is not definitively
clear, $n_+\la n_-$. Although the observational data are consistent
with annihilation in (hydrogen) plasma at rest, the redshift of the
line suggests that it could be also a stream of pair plasma with
matter captured from the source environment.

\begin{acknowledgements}
The authors thank Anatoly Tur for helpful discussions, useful comments
of an anonymous referee are greatly acknowledged. I.M.\ is grateful to the
{\it SIGMA} team of CESR for warm hospitality and facilities which made
his stay in Toulouse very fruitful and pleasant. This work was partly
supported through a post-doctoral fellowship of the French Ministry of
Research.

\end{acknowledgements}

\end{document}